\documentclass[article]{JHEP}
\usepackage{epsf,epsfig,amsfonts,amsmath,amssymb}

\newdimen\tableauside\tableauside=1.0ex
\newdimen\tableaurule\tableaurule=0.4pt
\newdimen\tableaustep
\def\phantomhrule#1{\hbox{\vbox to0pt{\hrule height\tableaurule width#1\vss}}}
\def\phantomvrule#1{\vbox{\hbox to0pt{\vrule width\tableaurule height#1\hss}}}
\def\sqr{\vbox{%
  \phantomhrule\tableaustep
  \hbox{\phantomvrule\tableaustep\kern\tableaustep\phantomvrule\tableaustep}%
  \hbox{\vbox{\phantomhrule\tableauside}\kern-\tableaurule}}}
\def\squares#1{\hbox{\count0=#1\noindent\loop\sqr
  \advance\count0 by-1 \ifnum\count0>0\repeat}}
\def\tableau#1{\vcenter{\offinterlineskip
  \tableaustep=\tableauside\advance\tableaustep by-\tableaurule
  \kern\normallineskip\hbox
    {\kern\normallineskip\vbox
      {\gettableau#1 0 }%
     \kern\normallineskip\kern\tableaurule}% 
  \kern\normallineskip\kern\tableaurule}}
\def\gettableau#1 {\ifnum#1=0\let\next=\null\else
  \squares{#1}\let\next=\gettableau\fi\next}

\newcommand{\twoVgraph}{\raisebox{0pt}{
                 \begin{picture}(18,18)(-9,-5)
                 \put(0,0){\circle{16}} \put(-8,0){\line(1,0){16}}
                 \end{picture}}}

\newcommand{\fourVgraph}{\raisebox{0pt}{
                 \begin{picture}(18,26)(-9,-9)
                 \put(0,0){\oval(16,24)} \put(-8,4){\line(1,0){16}}
                 \put(-8,-4){\line(1,0){16}}
                 \end{picture}}}

\newcommand{\sixVgraph}{\raisebox{0pt}{
                 \begin{picture}(26,24)(-13,-8)
                 \put(-9,8){\circle{6}} \put(9,8){\circle{6}}
                 \put(-6,8){\line(1,0){12}} \put(0,-8){\circle{6}} 
                 \put(-9,5){\line(2,-3){7}} \put(9,5){\line(-2,-3){7}}
                 \end{picture}}}

\newcommand{\eightVgraphI}{\raisebox{0pt}{
                 \begin{picture}(26,26)(-13,-9)
                 \put(-9,9){\circle{6}} \put(9,9){\circle{6}}
                 \put(-9,-9){\circle{6}} \put(9,-9){\circle{6}} 
	      	 \put(-6,9){\line(1,0){12}}
                 \put(-9,6){\line(0,-1){12}}
                 \put(-6,-9){\line(1,0){12}}
                 \put(9,6){\line(0,-1){12}}
                 \end{picture}}}

\newcommand{\eightVgraphII}{\raisebox{0pt}{
                 \begin{picture}(28,28)(-14,-10)
                 \put(-13,13){\line(1,0){26}}
                 \put(-13,-13){\line(1,0){26}}
                 \put(-13,-13){\line(0,1){26}}
                 \put(13,-13){\line(0,1){26}}
                 \put(-3,3){\line(1,0){6}}
                 \put(-3,-3){\line(1,0){6}}
                 \put(-3,-3){\line(0,1){6}}
                 \put(3,-3){\line(0,1){6}}
                 \put(-13,13){\line(1,-1){10}}
                 \put(-13,-13){\line(1,1){10}}
                 \put(13,13){\line(-1,-1){10}}
                 \put(13,-13){\line(-1,1){10}}
                 \end{picture}}}

\newcommand{\tenVgraphI}{\raisebox{0pt}{
                 \begin{picture}(26,29)(-13,-9)
                 \put(-9,9){\circle{6}} \put(9,9){\circle{6}}
                 \put(-9,-9){\circle{6}} \put(9,-9){\circle{6}}
                 \put(0,9){\circle{6}}
                 \put(-6,9){\line(1,0){3}}
                 \put(6,9){\line(-1,0){3}}
                 \put(-9,6){\line(0,-1){12}}
                 \put(-6,-9){\line(1,0){12}}
                 \put(9,6){\line(0,-1){12}}
                 \end{picture}}}

\newcommand{\tenVgraphII}{\raisebox{0pt}{
                 \begin{picture}(28,31)(-14,-10)
                 \put(-13,13){\line(1,0){10}}
                 \put(0,13){\circle{6}}
                 \put(13,13){\line(-1,0){10}}
                 \put(-13,-13){\line(1,0){26}}
                 \put(-13,-13){\line(0,1){26}}
                 \put(13,-13){\line(0,1){26}}
                 \put(-3,3){\line(1,0){6}}
                 \put(-3,-3){\line(1,0){6}}
                 \put(-3,-3){\line(0,1){6}}
                 \put(3,-3){\line(0,1){6}}
                 \put(-13,13){\line(1,-1){10}}
                 \put(-13,-13){\line(1,1){10}}
                 \put(13,13){\line(-1,-1){10}}
                 \put(13,-13){\line(-1,1){10}}
                 \end{picture}}}

\newsavebox{\DISK}
\savebox{\DISK}[8pt]{\begin{picture}(8,8)(0,0)
                     \put(-2.5,-3){$\bullet$}
                     \end{picture}}

\preprint{HUTP--02/A029\\  {\tt hep-th/0207096}}

\title{Chern-Simons theory, matrix integrals, \\ and perturbative 
three-manifold invariants}

\author{Marcos Mari\~no 
\\
Jefferson Physical Laboratory,\\
Harvard University,\\ Cambridge, MA 02138, USA\\
\email{marcos@born.harvard.edu}
}

\abstract{The universal
perturbative invariants of rational homology spheres can be 
extracted from the Chern-Simons 
partition function by combining perturbative and nonperturbative results. 
We spell out the 
general procedure to compute these invariants, and 
we work out in detail the case of Seifert spaces.     
By extending some previous results of Lawrence and Rozansky,  
the Chern-Simons partition function with arbitrary 
simply-laced group for these spaces is written 
in terms of matrix integrals. The analysis
of the perturbative expansion amounts to the
evaluation of averages in a Gaussian ensemble of random matrices. 
As a result, explicit expressions for the universal perturbative 
invariants 
of Seifert homology spheres up to order five are presented.
}


\begin{document}

%%%%%%%%%%%%%%%%%%%%%%%%%%%%%%%%%%%%%%%%%%%%%%%%%%%%%%%%%%%%%%%%%
%%%%%%%%%%%%%%%%%%%%%%%%%%%%%%%%%%%%%%%%%%%%%%%%%%%%%%%%%%%%%%%%%

%%%%%%%%%%%%%%%%%%%%%%%%%%%%%%%%%%%%%%%%%%%%%%%%%%%%%%%%%%%%%%%%%
%%%%%%%%%%%%%%%%%%%%%%%%%%%%%%%%%%%%%%%%%%%%%%%%%%%%%%%%%%%%%%%%%

\section{Introduction} \label{s0}

Chern-Simons theory \cite{cs} has been at the heart of the developments in
three-manifold topology and knot theory for the last ten years. The
partition function of Chern-Simons theory defines a topological 
invariant of three-manifolds, sometimes known as the
Witten-Reshetikhin-Turaev invariant, that can be studied from many
different points of view. In general, the invariant thus obtained contains
information about the three-manifold itself but also about the gauge theory
group that one uses to define the theory. 

However, from a perturbative point of view it is clear that one can extract 
numerical invariants of the three-manifold which are intrinsic to it 
and do not depend on the 
gauge group. This goes as follows: if we compute the partition function 
in perturbation theory, the contribution 
at a given order consists of a sum of terms associated 
to Feynman diagrams. Each term is the product of a group dependent factor (the
group weight of the diagram), and a factor involving multiple 
integrals of the propagators over the three-manifold. 
This last factor does not depend on the 
gauge group one started with, and in this sense it is universal. Therefore,
one can extract from perturbation theory an infinite series of
invariants, the so-called universal perturbative invariants of
three-manifolds. 

The idea of looking at the perturbative expansion of Chern-Simons theory in
order to extract numerical invariants that ``forget'' about the gauge group
was first implemented in the context of knot invariants, leading to the
theory of Vassiliev invariants and the Kontsevich integral (see
\cite{barnatanvass,labasrev}). The perturbative approach to the study of
the partition function of Chern-Simons theory 
has a long story, starting in \cite{cs}. This has been pursued from
many points of view. On the one hand, the structure of the perturbative
series has been analyzed in detail (see for example \cite{alr,as} and 
\cite{dij} for a nice review), leading to the graph
homology of trivalent graphs as a systematic tool to organize the
expansion. On the other hand, the asymptotic expansion of the
nonperturbative results has also been studied 
\cite{jeffrey,rozseifert, 
roz, rozone, lawrence,lr}, although so far all the analysis
have focused on theories with gauge group $SU(2)$. Finally, a
mathematically rigorous theory of universal perturbative invariants of
three-manifolds has been constructed starting from the Kontsevich
integral: the so-called LMO invariant \cite{lmo} and its Aarhus version
\cite{aarhus}. 
      
The main goal of the present paper is to elaborate on the topological 
field theory approach
to universal perturbative invariants. The point of view presented here 
is very similar to
the one advocated in \cite{al,alp} to extract Vassiliev invariants from
Chern-Simons perturbation theory: first, one analyzes the structure of the 
perturbative series of an observable in the theory. 
This means in practical terms choosing a basis of
independent group factors and compute its value for various gauge groups. 
In a second step, one computes the corresponding 
invariant nonperturbatively for
those gauge groups, 
performs an asymptotic
expansion, and extracts the universal invariants
 by comparing to the perturbative
result. This program was applied successfully in \cite{al,alp} to
compute Vassiliev invariants of many knots. It turns out that, in the case
of the Chern-Simons partition function, the first step is relatively easy,
but the calculation of the partition function for arbitrary gauge groups in
a way that is suitable for an asymptotic expansion turns out to be
trickier, except in very simple cases. 

In this paper, some well-known results concerning the structure of
the perturbative series are put together, 
and we carry out a detailed analysis up to order
five. The focus is 
on a rather general class of rational homology spheres, 
Seifert spaces. The partition function of Chern-Simons theory with gauge
group $SU(2)$ on these spaces and its asymptotic expansion have been 
studied in 
\cite{fg,rozseifert, rozone}. The extension to higher 
rank gauge groups has also been
considered \cite{takata,hansen}, but in forms that are not useful for
a systematic perturbative expansion. In \cite{lr},
Lawrence and Rozansky found a
beautiful expression for the $SU(2)$ partition function on Seifert spaces
in terms of a sum of integrals and residues. It turns out that their result 
can be generalized to any
simply-laced group and written in terms of integrals over the Cartan
subalgebra of the gauge group (these kind of integrals already appeared in
a related context in \cite{roz}). Interestingly, they are closely
related to models of random matrices, and one can use 
matrix model technology to study the Chern-Simons partition function on these
spaces. 
The resulting expressions can be expanded in series in a fairly systematic 
way, and by
comparing the result with the general structure of the perturbative
expansion, the universal perturbative invariants can be extracted. It should be mentioned that the full  
LMO invariant of Seifert spaces 
has been computed by Bar-Natan and Lawrence \cite{bl} by using techniques
from the theory of the 
Aarhus integral. However, their result is rather implicit and
involves a complicated graphical calculus. 

This paper is organized as follows: in section 2, we review the computation
of the Chern-Simons partition function starting from a surgery
presentation. In section 3, we analyze in some detail the structure of the
Chern-Simons perturbation series. In section 4 we compute the exact partition function of Seifert
spaces for simply-laced gauge groups, 
generalizing the results of Lawrence and Rozansky, and we make the
connection to matrix models. In section 5, we analyze the asymptotic
expansion of the exact result, explain how to evaluate the matrix
integrals, and present the results for universal perturbative invariants 
up to order five. In section 6, we comment on the possible relevance of
these results to other physical contexts, and some avenues for
future research are suggested. The Appendix collects the explicit
expressions for the group factors and the matrix integrals, together with a
summary of the properties of symmetric functions that are used in the paper.

\section{The partition function of Chern-Simons theory}

In this section we review some well-known results about the computation
of the Chern-Simons partition function in terms of surgery presentations. 
An excellent summary, that we follow quite closely, is given in \cite{roz}.

We consider Chern-Simons theory on a three-manifold $M$ and for a
simply-laced 
gauge group $G$, with action
\begin{equation}
S(A)={k \over 4 \pi} \int_M {\rm Tr} \Bigl( A \wedge dA + {2 \over 3} A\wedge
A 
\wedge A \Bigr),
\end{equation}
where $A$ is a $G$-connection on $M$. We will be interested in 
{\it framed} three-manifolds, {\it i.e.} a three-manifold together with a
trivialization of the bundle $TM \oplus TM$. As explained in \cite{atiyah},
for every three-manifold there is a canonical choice of framing, and the
different choices are labeled by an integer $s \in {\bf Z}$ in such a way
that $s=0$ corresponds to the canonical framing. Unless otherwise stated,
we will always work in the canonical framing, and we will explain below how
to incorporate this in the calculations, following \cite{jeffrey,fg,lr}. 

As shown by Witten in \cite{cs}, the partition function of Chern-Simons
theory
\begin{equation}
\label{wrt}
Z_k(M)= \int {\cal D} A {\rm e}^{i S_{\rm CS}(A)}.
\end{equation}
defines an invariant of framed manifolds. There is a very 
nice procedure to evaluate (\ref{wrt}) in a combinatorial way which
 goes as follows. By Lickorish theorem (see for example \cite{lick}), 
any three-manifold 
$M$ can be obtained by surgery on a link ${\cal L}$ in ${\bf S}^3$. 
Let us denote by 
${\cal K}_i$, $i=1, \cdots, L$, the components of ${\cal L}$. 
The surgery operation means that 
around each of the knots ${\cal K}_i$ we take a tubular
neighborhood ${\rm Tub}({\cal K}_i)$ that we remove from ${\bf S}^3$. This
tubular neighborhood is a solid torus with a contractible cycle $\alpha_i$ and
a noncontractible cycle $\beta_i$. We then glue the solid torus back after
performing an
${\rm SL}(2, {\bf Z})$ transformation given by the matrix
\begin{equation}
\label{sltwo}
U^{(p_i,q_i)}=\left( \begin{array} {cc} p_i & r_i \\ 
q_i & s_i\end{array} \right).
\end{equation}
This means that the cycles $p_i \alpha_i + q_i \beta_i$ and 
$r_i \alpha_i + s_i \beta_i$ on the boundary of the complement of ${\cal
K}_i$ are identified with the cycles $\alpha_i$, $\beta_i$ in ${\rm
Tub}({\cal K}_i)$. 

This geometric
description leads to the following prescription to compute the invariants
in Chern-Simons theory. By canonical quantization, one associates a Hilbert
space to any two-dimensional compact manifold that arises as the boundary
of a three-manifold, so that the path-integral over a manifold with
boundary gives a state in the corresponding Hilbert space. As it was shown
in \cite{cs}, the states of the Hilbert space of 
Chern-Simons theory associated to the torus are in one 
to one correspondence with 
the integrable representations of the WZW model with gauge group $G$ at
level $k$. We will use the following notations in the following: 
$r$ denotes 
the rank of $G$, and $d$ its dimension. 
$y$ denotes the dual Coxeter number. 
The fundamental weights will be denoted by
$\lambda_i$, and the simple roots by $\alpha_i$, with $i=1, \cdots, r$. 
The weight and root lattices of $G$ are
denoted by $\Lambda_{\rm w}$ and $\Lambda_{\rm r}$, respectively. 
Finally, we put $l=k +y$. 

A representation given by a highest weight $\Lambda$ is integrable if 
the weight $\rho + \Lambda$ is in the   
fundamental chamber ${\cal F}_l$ ($\rho$ denotes as usual the Weyl vector,
given by the sum of the fundamental weights). The fundamental chamber 
is given by $\Lambda_{\rm w}/l 
\Lambda_{\rm r}$ modded out by the action of the Weyl group. For example,
in $SU(N)$ a 
weight $p=\sum_{i=1}^r p_i \lambda_i$ is in ${\cal F}_l$ if  
\begin{equation}
\sum_{i=1}^r p_i < l,\,\,\,\,\,\, {\rm and} \,\,\ p_i >0, \, i=1, \cdots, r.
\end{equation}
In the following, 
the basis of integrable representations will be labeled by the 
weights in ${\cal F}_l$.  

In the case of simply-laced gauge groups, 
the ${\rm Sl}(2, {\bf Z})$ transformation given by $U^{(p,q)}$ has
the following matrix elements in the above basis \cite{jeffrey,roz}:
\begin{eqnarray}
{\cal U}^{(p, q)}_{\alpha \beta} &=& 
{[i \, {\rm sign}(q)]^{|\Delta_+|} \over (l |q|)^{r/2}}
\exp \Bigl[ -{ i d \pi \over 12}  \Phi (U^{(p,q)})\Bigr] 
\Biggl( { {\rm Vol}\, \Lambda_{\rm w} \over {\rm Vol}\,  \Lambda_{\rm r}} 
\Biggr)^{1 \over 2} \nonumber\\
& \cdot& \sum_{n \in \Lambda_{\rm r}/q \Lambda_{\rm r}}\sum_{w \in {\cal W}} \epsilon
(w) \exp \Bigl\{ {i \pi \over l q} ( p \alpha^2 - 2\alpha (l n + w(\beta))
+ s(ln + w(\beta))^2 \Bigr\}.
\end{eqnarray}
In this equation, $|\Delta_+|$ denotes the number of positive roots of $G$, 
and the second sum is over the Weyl group ${\cal W}$ of $G$.
 $\Phi (U^{(p,q)})$ is the Rademacher function:
\begin{equation}
\label{rade}
\Phi\left[ \begin{array}{cc} p & r \\ q&s\end{array} \right]= 
{p + s \over q} - 12 s(p,q),
\end{equation}
where $s(p,q)$ is the Dedekind sum
\begin{equation}
s(p,q)={1 \over 4q} \sum_{n=1}^{q-1} \cot \Bigl( {\pi n \over q}\Bigr) 
\cot \Bigl( {\pi n p\over q}\Bigr).
\end{equation} 

With these data we can already present Witten's result for the Chern-Simons
partition function of $M$. As before, suppose that $M$ is obtained by
surgery on a link ${\cal L}$ in ${\bf S}^3$. Then, the partition function
of $M$ is given by:
\begin{equation}
\label{partf}
Z(M, l)= {\rm e}^{i\phi_{\rm fr}} \sum_{\alpha_1, \cdots, \alpha_L
 \in {\cal F}_l}Z_{\alpha_1, \cdots, \alpha_L}({\cal L})\,  {\cal U}^{(p_1,
 q_1)}_{\alpha_1 \rho} \cdots {\cal U}^{(p_L,
 q_L)}_{\alpha_L \rho}.
\end{equation} 
In this equation, $Z_{\alpha_1, \cdots, \alpha_L}({\cal L})$ is the
invariant of the link ${\cal L}$ with representation $\alpha_i-\rho$ 
attached to its $i$-th component (recall that the weights in ${\cal F}_l$
are of the form $\rho + \Lambda$). The phase factor ${\rm e}^{i\phi_{\rm
fr}}$ is a framing correction that guarantees that the resulting invariant
is in the canonical framing for the three-manifold $M$. Its explicit
expression is:
\begin{equation}
\label{framcorr}
\phi_{\rm fr}= {\pi k d \over 12 l }\biggl(\sum_{i=1}^L \Phi(U^{(p_i,
q_i)}) - 3 \sigma \, ({\cal L})\biggr),
\end{equation}    
where $\sigma({\cal L})$ is the signature of the linking matrix of ${\cal
L}$.

\section{Chern-Simons perturbation theory}

The expression (\ref{partf}) gives the nonperturbative result for the 
partition function of $M$, and allows an explicit evaluation for many 
three-manifolds for any gauge group $G$ and level $k$. However, from the 
point of view of Chern-Simons perturbation theory, the partition function
can be also understood as an asymptotic series in $l^{-1}$, whose
coefficients can be computed by evaluating Feynman diagrams. 
In this section we review some known facts about the perturbative expansion 
of Chern-Simons theory and we state our strategy to compute the
universal perturbative invariants.

We are interested in the perturbative 
evaluation of the partition function (\ref{wrt}).
Let us assume (as we will do in this paper) that 
$M$ is a rational homology sphere. The classical solutions of the
Chern-Simons action are just flat connections on $M$, and for a rational 
homology sphere these are a finite set of points. Therefore, in the
perturbative evaluation one expresses $Z_k(M)$ as a
sum 
of terms associated to stationary points:
\begin{equation}
Z_k (M)= \sum_c Z_k^{(c)}(M),
\end{equation}
where $c$ labels the different flat connections $A^{(c)}$ on $M$. 
Each of the terms in this sum has a perturbative expansion as an
asymptotic series in $l^{-1}$. 
The structure of the perturbative series was analyzed in
various papers \cite{cs,rozone,as} and is given by the following
expression:
\begin{equation}
\label{perts}
Z_k^{(c)}(M)=Z^{(c)}_{\rm 1-loop}(M). \exp \Biggl\{ \sum_{\ell=1}^\infty 
S^{(c)}_\ell  x^\ell 
\Biggr\}.
\end{equation}
In this equation, $x$ is the effective expansion parameter:
\begin{equation}
\label{coupling}
x = { 2 \pi i \over l}.
\end{equation} 
The one-loop correction $Z^{(c)}_{\rm 1-loop}(M)$ was first analyzed in
\cite{cs}, and has been studied in big detail since then. It has the form,
\begin{equation}
\label{asympt}
Z^{(c)}_{\rm 1-loop}(M)= 
{ (2 \pi x)^{ {1\over 2}({\rm dim}H^0_c - {\rm dim}H^1_c 
)} 
\over {\rm vol} (H_c)} {\rm e}^{-{1 \over x} S_{\rm CS}(A^{(c)}) 
-{i \pi \over 4}\varphi} {\sqrt {| \tau^{(c)}_R|}},
\end{equation}
where $H^{0,1}_c$ are the cohomology groups with values in the Lie algebra
of $G$ associated to the flat connection $A^{(c)}$, $ \tau^{(c)}_R$ is the 
Reidemeister-Ray-Singer torsion of $A^{(c)}$, $H_c$ is the
isotropy group of $A^{(c)}$, and $\varphi$ is a certain phase. More details about the
structure 
of this term can be found in \cite{cs,fg,jeffrey,rozseifert, roz}. 

Our main 
object of concern in this paper are the terms in the exponential of 
(\ref{perts}) corresponding to the {\it trivial} connection, which we
will 
simply denote by $S_{\ell}$. In order to make a precise statement about the 
structure of these terms, we have to 
explain in some detail what is the 
appropriate set of diagrams we want to consider. In principle, in order to
compute $S_\ell$ we just have to consider 
all the connected bubble diagrams with
$\ell$ loops. To each of these diagrams we 
will associate a group factor times  
a Feynman integral. However, not all these diagrams are independent, 
since the underlying Lie algebra structure imposes the Jacobi identity:
\begin{equation}
\label{ihx}
\sum_e \bigl( f_{abe}f_{edc} + f_{dae} f_{ebc} + f_{ace} f_{edb}\bigr)=0.    
\end{equation}
This leads to the diagram relation known as 
IHX relation. Also, antisymmetry of $f_{abc}$ leads to the so-called 
AS relation (see for example 
\cite{barnatanvass,dij,labasrev,sawth}). 
The existence of these relations between diagrams 
suggests to define an equivalence relation in the space
of connected trivalent graphs by quotienting by the IHX and the AS
relations, and this gives the so-called {\it graph homology}. The
space 
of homology classes of connected diagrams will be denoted by ${\cal
A}(\emptyset)^{\rm conn}$. This space is graded by half the number of
vertices, and this number gives the degree of the graph. The space of homology
classes of graphs at degree $\ell$ is then denoted by ${\cal
A}(\emptyset)^{\rm conn}_\ell$. For every $\ell$, this is a
finite-dimensional vector space of dimension $d(\ell)$. The dimensions of
these spaces are explicitly known for low degrees (see for example
\cite{sawth}), and we have listed some 
of them in Table \ref{dims}. 
Finally, notice that, 
given any group $G$, 
we have a map
\begin{equation}
{\cal A}(\emptyset)^{\rm conn} \longrightarrow {\bf R}
\end{equation}
that associates to every graph $\Gamma$ its group theory 
factor $r_{\Gamma}(G)$. This map is an example of a {\it weight system}
for ${\cal A}(\emptyset)^{\rm conn}$. Every gauge group gives a weight
system for ${\cal A}(\emptyset)^{\rm conn}$, but one may in principle find
weight systems not associated to gauge groups, although so far the only
known example is the one constructed by Rozansky and Witten in \cite{rw},
which uses instead hyperK\"ahler manifolds.
\begin{table}[htbp] 
\centering 
\begin{tabular}{|c|c|c|c|c|c|c|c|c|c|c|}\hline
 $\ell$ & 1 & 2 & 3 & 4 & 5 & 6 & 7 & 8 & 9 & 10  \\  \hline
 $d(\ell)$ & 1 & 1 & 1 & 2 & 2 & 3 & 4 & 5 & 6 & 8  \\ 
\hline
\end{tabular}
\caption{Dimensions $d(\ell)$ of ${\cal A}(\emptyset)^{\rm conn}_\ell$ up
to $\ell=10$.}
\label{dims}
\end{table}

We can now state very precisely what is the structure of the $S_\ell$
appearing in (\ref{perts}): since the Feynman diagrams can be grouped into 
homology classes, we have
\begin{equation}
\label{highloops}
S_\ell = \sum_{\Gamma \in 
{\cal A}(\emptyset)^{\rm conn}_\ell} r_{\Gamma}(G) I_{\Gamma}(M).
\end{equation}
The factors $ I_{\Gamma}(M)$ appearing in (\ref{highloops}) are certain
(complicated) integrals 
of propagators over $M$. It was shown in \cite{as} that these are 
differentiable invariants of the three-manifold $M$, and since the 
dependence on the gauge group has been factored out, they only capture 
topological information of $M$, in contrast to $Z_k(M)$, which also depends 
on the choice of the gauge group. These are the {\it universal perturbative
invariants} defined by Chern-Simons theory. Notice that, at every order
$\ell$ in perturbation theory, there are
$d(\ell)$ independent perturbative 
invariants. Of course, these invariants inherit from 
${\cal A}(\emptyset)^{\rm conn}_\ell$ the structure of a 
finite-dimensional vector 
space, and it is convenient to pick a basis once and for all. Here 
we will study these invariants up to order $5$, and we choose the basis 
presented by Sawon in \cite{sawth}:
\begin{eqnarray}
\label{graphi}
\ell=1: & & \,\,\,\,\, \twoVgraph \nonumber\\
 \ell=2: & & \,\,\,\,\, \fourVgraph \nonumber\\
\ell=3: & & \,\,\,\,\, \sixVgraph \nonumber\\
\ell=4: & & \,\,\,\,\, \eightVgraphI \,\,\,\,\, 
\eightVgraphII \nonumber\\  
\ell=5: & & \,\,\,\,\,  \tenVgraphI \,\,\,\,\, 
\tenVgraphII \nonumber\\
\end{eqnarray}
As in \cite{sawth}, we will denote the graphs with $k$ circles 
joined by lines by $\theta_k$. Therefore, the graph corresponding to 
$\ell=1$ will be denoted by $\theta$, the graph corresponding to 
$\ell=2$ will be denoted $\theta_2$, and so on. The second graph for
$\ell=4$ will be denoted by $\omega$, and the second graph in $\ell=5$ by 
$\omega\theta$. The group 
factors associated to these diagrams can be easily 
computed by using the techniques of \cite{cvita} (see also 
\cite{barnatanth,barnatanvass}). Explicit results for 
all classical gauge groups are presented in the Appendix. 

\medskip
{\bf Remarks}: 

\medskip

1. It is interesting to understand the framing 
dependence of the universal perturbative invariants (see \cite{as} for a
discussion of this issue). 
As shown in \cite{cs}, the full partition theory $Z_k(M)$ changes 
as follows under a change of framing:
\begin{equation}
Z \rightarrow {\rm e}^{\pi i s c\over 12} Z,
\end{equation}
where $s\in {\bf Z}$ labels the choice of framing and 
\begin{equation}
c={k d \over k+y}
\end{equation}
is the central charge of the WZW model with group $G$. Using now 
that (see Appendix A)
\begin{equation}
r_{\theta}(G)= 2 y d,
\end{equation}
we find that under a change of framing one has
\begin{equation}
I_{\theta}(M) \rightarrow I_{\theta}(M) - {s \over 48},
\end{equation}
while the other universal perturbative invariants remain the same. Since we
will work in the canonical framing of $M$, this will produce a canonical
value of $I_{\theta}(M)$. 

2. Notice that Chern-Simons theory detects the graph 
homology through the weight
   system associated to Lie algebras. Unfortunately it is known
   \cite{vogel} that there is an element of graph homology at 
   degree $16$ that it is not detected by any weight system associated to 
simple Lie algebras. However, there is a 
very elegant mathematical definition of the
   universal perturbative invariant of a three-manifold 
 that works directly in the graph
   homology. This is called the LMO invariant \cite{lmo} and it is a formal 
linear combination of homology graphs with rational coefficients: 
\begin{equation}
\label{lmoinv}
\omega (M) =\sum_{\Gamma \in {\cal A}(\emptyset)^{\rm conn}}
I^{\rm LMO}_{\Gamma}(M)\, \Gamma 
\in {\cal A}(\emptyset)^{\rm conn}[{\bf Q}]. 
\end{equation}  
It is believed that the universal invariants extracted from 
Chern-Simons perturbation theory agree with the LMO invariant. More
precisely, 
since the LMO invariant $\omega (M)$ is taken to 
be $0$ for ${\bf S}^3$, we have:
\begin{equation}
I^{\rm LMO}_{\Gamma}(M)=I_{\Gamma}(M)-I_{\Gamma}({\bf S}^3),
\end{equation}
as long as the graph $\Gamma$ is detected by Lie algebra weight systems. 
In that sense the LMO invariant is more refined than the universal
perturbative invariants extracted from 
Chern-Simons theory.

3. The Chern-Simons approach to the theory of universal 
perturbative invariants is very similar 
to the approach to Vassiliev invariants based on the analysis 
of vevs of Wilson loops in perturbation 
theory \cite{al,alp}. The role of graph homology is 
played there by the homology of chord diagrams (see 
for example \cite{barnatanvass,labasrev}).

\section{The Chern-Simons partition function on Seifert spaces}

In this section we write the partition function of Chern-Simons theory 
on Seifert homology spheres as a sum of integrals over the Cartan 
subalgebra and
a set of residues, by extending results of Lawrence and Rozansky \cite{lr} 
for $SU(2)$. We also show that these integrals can be interpreted in 
terms of matrix
integrals associated to a random matrix model.

\subsection{Seifert homology spheres}

Seifert homology spheres
 can be constructed by performing surgery on a 
link ${\cal L}$ in ${\bf S}^3$ 
with $n+1$ components, consisting on $n$ parallel and unlinked unknots 
together with a single unknot whose linking number with each of the other
$n$ unknots is one. The surgery data are $p_j/q_j$ for the unlinked 
unknots, $j=1, \cdots, n$, and 0 on the final component. $p_j$ is coprime
 to $q_j$ for all $j=1, \cdots, n$, and the $p_j$'s are pairwise coprime. 
After doing 
surgery, one obtains the Seifert space 
$M=X({p_1 \over q_1}, \cdots, {p_n
\over q_n})$. This is rational homology sphere whose 
first homology group $H_1(M, {\bf Z})$ has order $|H|$, where
\begin{equation}
\label{orderh}
H=P \sum_{j=1}^n {q_j \over p_j}, \,\,\,\,\, {\rm and} 
\,\,\,\,  
P=\prod_{j=1}^n p_j.
\end{equation}
Another topological invariant that will enter the computation is the 
signature of ${\cal L}$, which turns out to be \cite{lr}
\begin{equation}
\label{signa}
\sigma ({\cal L})=\sum_{i=1}^n {\rm sign}\biggl( {q_i \over p_i} \biggr) 
-{\rm sign}\biggl( {H \over P} \biggr).
\end{equation}
For $n=1,2$, Seifert homology spheres reduce to lens
spaces, and one has that $L(p,q)=X(q/p)$. For
$n=3$, we obtain the Brieskorn homology spheres $\Sigma (p_1, p_2, p_3)$
(in this case the manifold is independent of $q_1, q_2 ,q_3$). In
particular, $\Sigma (2,3,5)$ is the Poincar\'e homology sphere. Finally, 
the Seifert manifold $X({2\over -1}, {m\over (m+1)/2}, {t-m \over 1})$,
with $m$ odd,  can
be obtained by integer surgery on a $(2, m)$ torus knot with framing
$t$. 

\subsection{Computation of the partition function}

In order to compute the partition function of $M$, we first have to 
compute the invariant of ${\cal L}$ for generic representations
$\beta-\rho, \Lambda_1, \cdots, \Lambda_n$ of the gauge group $G$, where 
$\beta-\rho$ is the irreducible representation coloring the unknot with 
surgery data $0$, and $\Lambda_i$ are irreducible representations coloring 
the unknots with surgery data $p_i/q_i$, $i=1,\cdots, n$. This can be 
easily done by using
the formula of \cite{cs} for connected sums of knots, and one obtains: 
\begin{equation}
\label{qginvs}
Z_{\beta,\rho+ \Lambda_1, \cdots, \rho + \Lambda_n}({\cal L})=  
{\prod_{i=1}^n S_{\beta \rho + \Lambda_i} \over S_{\rho \beta}^{n-1}}.   
\end{equation}
Therefore, the partition function of $M$ will be given by
\begin{equation}
\label{partseif}
Z_k(M)={\rm e}^{i \phi_{\rm fr}}\sum_{\beta \in {\cal F}_{l}} 
{ \prod_{i=1}^n \sum_{\rho +\Lambda_i \in {\cal F}_{l}} S_{\beta 
\rho+ \Lambda_i}
{\cal U}^{(p_i, q_i)}_{\rho +\Lambda_i \, \rho} \over S_{\rho \beta}^{n-2}},
\end{equation}
where the framing correction is given by the general formula
(\ref{framcorr}). Seifert homology spheres can be also obtained by doing surgery on
$n$ strands parallel to ${\bf S}^1$ in ${\bf S}^2 \times {\bf S}^1$
\cite{rozseifert}, and then (\ref{partseif}) follows from 
Verlinde's formula \cite{verlinde}.

This expression is not suitable for an asymptotic expansion in $1/l$, since 
it involves a sum over integrable representations that depends itself on
$l$. In order to obtain a useful expression, we follow a series of steps
generalizing the procedure in \cite{roz,lr}. First of all, we perform the
matrix multiplication $\sum_{\rho + \Lambda_i\in {\cal F}_{l}} 
S_{\beta \rho + \Lambda_i}
{\cal U}^{(p_i, q_i)}_{\rho + \Lambda_i \, \rho}$. This gives
\begin{equation}
\sum_{\rho + \Lambda_i\in {\cal F}_{l}}S_{\beta \rho + \Lambda_i}
{\cal U}^{(p_i, q_i)}_{\rho + \Lambda_i \, \rho} = 
\exp \biggl( {\pi i k d \over 4 l } 
{\rm sign}\Bigl( {q_i \over p_i} \Bigr) \biggr) 
{\cal U}^{(-q_i, p_i)}_{\beta \rho},
\end{equation}
where the ${\rm SL}(2, {\bf Z})$ transformation in the right hand side is
given by
\begin{equation}
S \cdot U^{(p_i, q_i)}=\left( \begin{array} {cc} -q_i & -s_i \\ 
p_i & r_i\end{array} \right)=U^{(-q_i, p_i)},
\end{equation}
and the phase factor is needed in order to keep track of the framing. 
The partition function is then, up to a multiplicative constant, given by:
\begin{eqnarray}
\label{summ}
& & 
\sum_{\beta \in {\cal F}_l} {1 \over \prod_{\alpha >0} 
\Bigl( \sin {\pi \over
l} (\beta \cdot \alpha)  \Bigr)^{n-2}} \nonumber\\
& & \prod_{i=1}^n \sum_{n_i \in
\Lambda_{\rm r}/p_i \Lambda_{\rm r}} \sum_{w_i \in {\cal W}} \epsilon (w_i) 
\exp \biggl\{ {i \pi \over lp_i} \bigl( -q_i \beta^2 - 2 \beta ( l n_i + 
w(\rho)) + r_i (l n_i + w(\rho))^2\bigr) \biggr\}.
\end{eqnarray}
If $G$ is simply-laced, the summand is invariant under the simultaneous shift,
\begin{equation}
\label{shione}
 \beta \rightarrow \beta + l \alpha, \,\,\,\,\,\,\, 
n_i \rightarrow n_i - q_i \alpha,
\end{equation}
and also under
\begin{equation}
\label{shitwo}
n_i \rightarrow n_i + p_i \alpha.
\end{equation}
In these equations, $\alpha$ is any element in the root lattice. 
This invariance allows us put $n_i=0$ in
the above sum by extending the range of $\beta$: $\beta= p + l \alpha$, 
where $p \in {\cal F}_l$, and $\alpha= \sum_i a_i \alpha_i$, $
 0\le a_i<P$. It is easy to see 
that the resulting summand is invariant under the
Weyl group ${\cal W}$ 
acting on $\beta$, and by translations by $lP \alpha$, where $\alpha $ is
any root. We can then sum over Weyl reflections and 
divide by the order of ${\cal W}$, denoted by $|{\cal W}|$, 
and use the translation symmetry 
to extend the sum
over $\beta$ in the above set to a sum over $\beta \in (\Lambda_{\rm
w}/l P \Lambda_{\rm r})\backslash {\cal M}$. Here ${\cal M}$ denotes 
the set given by the wall of ${\cal F}_l$ together with 
its Weyl reflections and translations by $lP \alpha$ inside $\Lambda_{\rm
w}/l P \Lambda_{\rm r}$ (for $SU(N)$, the wall of ${\cal
F}_l$ is given by the weights with $\sum_i p_i =l$). We won't need a precise
description of the points of ${\cal M}$ in 
the following, since they only enter in the
contribution of irreducible flat connections to the path integral
\cite{lr}. 
After performing 
all these changes, and using the Weyl denominator formula   
\begin{equation}
\label{wden}
\prod_{\alpha>0 } 2 \sinh {\alpha \over 2} = 
\sum_{w \in {\cal W}}\epsilon(w) {\rm
e}^{w(\rho)},
\end{equation}
we can write (\ref{summ}) as:
\begin{eqnarray}
\label{almostfin}
& & {1 \over | {\cal W}|} 
{\rm e}^{ {i \pi  \over l}\rho^2 \sum_{i=1}^n { r_i\over p_i}} 
\sum_{\beta \in (\Lambda_{\rm
w}/lP \Lambda_{\rm r}) \backslash {\cal M}}
 {1 \over \prod_{\alpha >0} 
\Bigl( \sin {\pi \over
l} (\beta \cdot \alpha) \Bigr)^{n-2}}\nonumber\\
& &\cdot {\rm e}^{-{i \pi H \over lP}\beta^2} \prod_{i=1}^n \prod_{\alpha>0} 
(-2i) \sin {\pi \over
l p_i } (\beta \cdot \alpha).
\end{eqnarray}
The last step involves transforming the above sum in a sum over integrals 
and residues. To do that, we generalize slightly \cite{lr} and we introduce
a holomorphic function of $\beta_1, \cdots, \beta_r$ and $x_1, \cdots, x_r$
given by:
\begin{equation}
\label{aga}
h (\beta, x)= {{\rm e}^{-{i \pi H \over lP}\beta^2} \over 
 \prod_{\alpha >0} 
\Bigl( {\rm e}^{{\pi i \over
l} (\beta \cdot \alpha)} - {\rm e}^{-{\pi i \over
l} (\beta \cdot \alpha)}\Bigr)^{n-2}} 
{{\rm e}^{{2 \pi i\over l} \beta\cdot x } \over 
 \prod_{i=1}^r (1- {\rm e}^{-2 \pi i \beta_i})} = {f (\beta, x) 
\over \prod_{i=1}^r (1- {\rm e}^{-2 \pi i \beta_i})},
\end{equation}
where $\beta= \sum_{i=1}^r \beta_i \lambda_i \in \Lambda_{\rm w} 
\otimes {\bf C}$, $x= \sum_{i=1}^r x_i \alpha_i \in \Lambda_{\rm r} 
\otimes {\bf C}$. This function satisfies:
\begin{equation}
\label{period}
h (\beta+ lP \alpha, x) = {\rm e}^{2 \pi i P \alpha \cdot x} h (\beta, 
x-lH \alpha),
\end{equation}
for any $\alpha \in \Lambda_{\rm r}$. Notice also that $h (\beta, x)$ has 
poles at the points of $\Lambda_{\rm w}$, the weight lattice. 
Introduce now the integral
over ${\bf C}^r$:
\begin{equation}
\Theta (x)=\int_{C^r} h (\beta, x) d\beta
\end{equation}
where $C^r=C \times\cdots \times C$ is a multiple contour in ${\bf C}^r$,
and $C$ is the contour considered in \cite{lr}: a line through the origin 
from $(-1+ i) \infty$ to $(1- i) \infty$ for ${\rm sign}(H/P)>0$ (if 
${\rm sign}(H/P)<0$, we rotate $C$ by $\pi/2$ in the clockwise
direction). This contour is chosen to guarantee good convergence properties
as $\beta_i \rightarrow \infty$. 

Let us 
now shift the contour in such a way that it crosses all the poles
corresponding to the weights in the
chamber $\Lambda_{\rm w}/lP\Lambda_{\rm r}$.    
Using (\ref{period}) it is easy to see that, if 
$P \alpha \cdot x \in {\bf Z}$ for any root $\alpha$, 
the resulting integral can be written as  
\begin{equation}
\label{multishift}
\sum_{i=1}^r \Theta(x-lH \alpha_i) -
\sum_{1 \le  i<j\le r} \Theta(x-lH  (\alpha_i + \alpha_j)) + \cdots + 
(-1)^{r-1} \Theta(x-lH \sum_{i=1}^r \alpha_i).
\end{equation}
The difference between the
original integral and the shifted integral (\ref{multishift}) 
can be written as 
\begin{equation}
\sum_{t \in \Lambda_{\rm r}/H 
\Lambda_{\rm r}} \int_{C^r} f(\beta,x) {\rm e}^{-2 \pi i t \cdot \beta}.
\end{equation}
On the other hand, the effect of shifting the contour is to pick 
the residues corresponding to all the weights in the
chamber $\Lambda_{\rm w}/lP\Lambda_{\rm r}$. Here the residue 
is understood as $\lim_{\beta_i \rightarrow n_i} 
\prod_i (\beta_i -n_i)h (\beta, x)$, and the residues for the
weights that are not in ${\cal M}$ are simply given by $(2 \pi i )^{-r}
f(\beta,x)$. 
Putting everything together we find,
\begin{equation}
\label{residua}
\sum_{n \in  (\Lambda_{\rm
w}/lP \Lambda_{\rm r}) \backslash  {\cal M}}
f(n, x)= \sum_{t \in \Lambda_{\rm r}/H\Lambda_{\rm r}}
\int_{C^r}  f(\beta, x) {\rm e}^{ -2 \pi i \beta \cdot t} \, d \beta 
- (2 \pi i)^r \sum_{n \in  {\cal M}}
{\rm Res}(h (\beta, x), \beta=n),
\end{equation}
whenever $P \alpha \cdot x \in {\bf Z}$. We can apply this formula to   
(\ref{almostfin}), since what we have there is just a sum of 
expressions of the form $f(\beta, x)$ in (\ref{aga}), with $x$ of the form 
$\alpha/P$, $\alpha \in \Lambda_{\rm r}$. In this context, the sum over 
$t \in \Lambda_{\rm r}/H\Lambda_{\rm r}$ is interpreted as a sum over
reducible flat connections on the Seifert sphere, and of course $t=0$
corresponds to the trivial connection. In the remaining of this paper we
will focus on these contributions, {\it i.e.} we will not deal with the 
residue terms in (\ref{residua}), that should give the contribution of 
irreducible flat connections \cite{rozone,lr}. In fact, we will only
analyze in detail the contribution of the trivial connection in order to
make contact with the universal perturbative invariants. 

In order to present the final result for the contribution of reducible
flat connections to the partition function of Chern-Simons theory on
Seifert spaces, we have to collect the prefactors, including the
phases. Define as in \cite{lr}:
\begin{equation}
\label{fi}
\phi = 3\, {\rm sign} \Bigl( {H \over P} \Bigr) + \sum_{i=1}^n \Bigl( 12\,
s(q_i, p_i) - {q_i \over p_i} \Bigr).
\end{equation}
Therefore, the contribution of reducible flat connections to the
Chern-Simons partition function of $X({p_1 \over q_1}, \cdots, {p_n
\over q_n})$ is given by 
\begin{eqnarray}
\label{fians}
& & {(-1)^{|\Delta_+|} \over | {\cal W}|\, (2 \pi i )^r} 
\Biggl( { {\rm Vol}\, \Lambda_{\rm w} \over {\rm Vol}\,  \Lambda_{\rm r}} 
\Biggr){ [{\rm sign}(P)]^{|\Delta_+|} \over |P|^{r/2}} 
{\rm e}^{{ \pi i d \over 4} {\rm sign}(H/P) -{\pi i dy \over 12 l} \phi}
\nonumber\\
& \cdot & \sum_{t \in \Lambda_{\rm r}/H\Lambda_{\rm r}}
\int d\beta \,  {\rm e}^{ -{\beta^2/2 \hat x} - l t\cdot \beta}{ 
\prod_{i=1}^n \prod_{\alpha>0} 2 \sinh {\beta \cdot \alpha \over
2 p_i } \over \prod_{\alpha >0} 
\Bigl( 2 \sinh { \beta \cdot \alpha \over
2} \Bigr)^{n-2}}
\end{eqnarray}
In this equation, $\phi$ is given by (\ref{fi}), and in obtaining the phase 
factor we have made use of the Freudenthal-De Vries formula
\begin{equation}
\rho^2 ={1 \over 12} dy.
\end{equation}
We have also introduced the hatted coupling constant
\begin{equation}
\hat x = { Px \over H}, 
\end{equation}
where $x$ is the coupling constant given in (\ref{coupling}). In the
evaluation of the above integral we can rotate the integration 
contour $C^r$ to ${\bf R}^r$ as long as we are 
careful with phases in the Gaussian integral, as explained for example in \cite{cs}. If we
specialize (\ref{fians}) to 
$G=SU(2)$, we obtain the result derived in \cite{lr}. 
The expression 
(\ref{fians}) is in principle only valid for simply-laced groups, although
the results for the perturbative series turn out to be valid for any gauge
group. 

Notice that, in the sum over $\Lambda_{\rm r}/H\Lambda_{\rm r}$,
the $t$'s that are related by Weyl transformations 
correspond to the same flat connection. Fortunately, each of the 
integrals in (\ref{fians}) is
invariant under Weyl permutations of $t$, so in order to consider the
contribution of a given flat connection, one can just evaluate
(\ref{fians}) for a particular representative and then multiply by the
corresponding degeneracy factor ({\it i.e.} the number of Weyl-equivalent
$t$ configurations giving the same flat connection).   

If one is just interested in obtaining 
the contribution of the trivial
connection, one can use the shorter arguments of \cite{roz} and end up
with (\ref{fians}) with $t=0$. The contribution of the reducible
connections can also be obtained by generalizing the arguments of
\cite{rozone} to the higher rank situation. 
   
\subsection{Connection to matrix models}
In (\ref{fians}) we have written the contribution of reducible connections 
to the Chern-Simons partition function in terms of an integral over the 
Cartan subalgebra, since $d\beta =\prod_{i=1}^r d\beta_i$ and $\beta_i$ are
the Dynkin coordinates. In fact, the above expression can be interpreted as
the partition function of a random matrix model (for a review of random
matrices, see \cite{mehta,id}). To see this, let us consider a slight
generalization of the above results to the $U(N)$ and $O(2r)$
theories. The partition function for these groups 
can be obtained by writing $\beta$ in terms of the orthonormal basis in the space
of weights. 

Let us first consider the case of $U(N)$. Denote the orthonormal basis as
$\{ e_k \}_{k=1, \cdots, N}$, and put $\beta=\sum_k \beta_k e_k$ (where 
$\beta_k$ are taken to be independent variables), $t=\sum_k t_k e_k$. 
It is well-known that the
positive roots can be written as
\begin{equation}
\label{ubasis}
\alpha_{kl}=e_k - e_l,\,\,\,\,\,\,\, 1\le k<l\le N.
\end{equation}
Therefore, the integral in (\ref{fians}) becomes
\begin{equation}
\label{unmat}
\int d \beta \,  {\rm e}^{-\sum_k \beta_k^2 /2 \hat x - l\sum_k t_k \beta_k } 
{\prod_{i=1}^n \prod_{k<l} 2 \sinh {\beta_k -\beta_l\over
2 p_i } \over \prod_{k<l} 
\Bigl( 2 \sinh { \beta_k - \beta_l  \over
2} \Bigr)^{n-2}} 
\end{equation}
We can interpret the $\beta_k$ as the eigenvalues of a Hermitian matrix in
a Gaussian potential and interacting through
\begin{equation}
\label{unpote}
\sum_{i=1}^n \sum_{k<l} 
\log \biggl( 2 \sinh {\beta_k -\beta_l\over
2 p_i }\biggr) + (2-n)\sum_{k<l} 
\log \biggl( 2 \sinh {\beta_k -\beta_l\over
2 }\biggr).
\end{equation}
Notice moreover that for a small separation of the eigenvalues
(\ref{unpote}) becomes, at leading order,
\begin{equation}
\sum_{k<l} 
\log ( \beta_k -\beta_l )^2 
\end{equation}
which is the interaction between eigenvalues of the 
standard Hermitian matrix model. Therefore, the integral above can be
interpreted as a nonlinear deformation of the usual Gaussian unitary
ensemble (GUE). In fact, as we will see in detail in the next section, 
Chern-Simons perturbation theory means that 
we expand around the GUE, and the perturbative corrections are obtained by
evaluating averages in this ensemble. 
Note that the non-trivial reducible flat connections, labeled by $t$, are
interpreted in the matrix model language as a 
source term coupling linearly to the
eigenvalues. 

Similar considerations apply to the orthogonal group $O(2r)$. 
The positive roots can be 
written in terms of an orthonormal basis as follows:
\begin{equation}
\label{obasis}
\alpha_{kl}^{\pm}=e_k \pm e_l, \,\,\,\,\,\,\, 1\le k<l\le r,
\end{equation}
and the interaction between the eigenvalues reduces again, in the limit of 
small separation, to 
\begin{equation}
\label{goeint}  
\sum_{k<l} \log (\beta_k^2 - \beta_l^2)^2,
\end{equation}
which is the eigenvalue interaction of the orthogonal ensemble $O(N)$ for
even $N$.

\section{Asymptotic expansion and matrix integrals}

\subsection{Asymptotic expansion of the exact result}
In this subsection we will study the asymptotic expansion of the 
exact result obtained in the previous section for the contribution of the
trivial connection $t=0$. 

The expression (\ref{fians}) is very well suited for an asymptotic
expansion in powers of $x^\ell$: we just have to expand the integrand in a
power series of $\beta$, and integrate the result term by term with the
Gaussian weight. The integrand has the expansion:
\begin{equation}
\label{inte}
{ \prod_{i=1}^n \prod_{\alpha>0} 2 \sinh {\beta \cdot \alpha \over
2 p_i } \over \prod_{\alpha >0} 
\Bigl( 2 \sinh { \beta \cdot \alpha \over
2} \Bigr)^{n-2}} = {1 \over P^{|\Delta_+|}} \Bigl( \prod_{\alpha>0} 
(\beta \cdot \alpha)^2
\Bigr) f(\beta),
\end{equation}
where $f(\beta)$ has the form
\begin{equation}
\label{integrand}
f(\beta)= \prod_{\alpha>0} \biggl( 1 + \sum_{s=1}^\infty a_s
(\beta \cdot \alpha)^{2s}\biggr),\end{equation}
The coefficients $a_s$ can be obtained in a very straightforward way from  
(\ref{inte}). They are polynomials of degree $s$ in $n$ and in the power sums
\begin{equation}
\pi_j =\sum_{i=1}^n p_i^{-2j}.
\end{equation}
One has, for example, 
\begin{eqnarray}
a_1 &=& {1 \over 24} (\pi_1 + 2-n), \nonumber\\
a_2 & =& {1 \over 5760} ( 16 + 5n^2 -18 n -10 n \pi_1 + 20 \pi_1 + 
5\pi_1^2 -2 \pi_2).
\end{eqnarray}
Let us analyze in more detail the structure of $f(\beta)$. Define 
\begin{equation}
\sigma_j (\beta) =\sum_{\alpha>0} (\beta \cdot \alpha)^{2j}.
\end{equation}
By taking the log of (\ref{integrand}), one finds:
\begin{equation}
f(\beta)= \exp\biggl( \sum_{k=1}^\infty a_k^{(c)} \sigma_k (\beta)\biggr),
\end{equation} where the connected coefficients $a_k^{(c)}$ are defined in
the usual way: $\log (1 + \sum_n a_k x^k)= \sum_k a_k ^{(c)} x^k$. An
explicit expression for $f(\beta)$ can be obtained as follows. Let $\vec k
=(k_1, k_2, \cdots)$ be a vector whose components are nonnegative integers. 
Denote $\ell =\sum_j j k_j$, and define:
\begin{equation}
a_{\vec k}^{(c)}= \prod_j (a_j ^{(c)})^{k_j}, \,\,\,\,\,\,\,\,\,\,\, 
\sigma_{\vec k}(\beta)=\prod_j \sigma_j^{k_j} (\beta).
\end{equation}
Then, 
\begin{equation}
f(\beta) = 1 + \sum_{\vec k}{1 \over {\vec k}!} 
a_{\vec k}^{(c)}
 \sigma_{\vec k}(\beta),
\end{equation}
with ${\vec k}! = \prod_j k_j!$ and the sum is over all 
vectors $\vec k$. 
We see that the perturbative expansion of the partition function 
can be written in terms of the quantities
\begin{equation}
\label{matint}
{\cal R}_{\vec k}(G) = {\int d \beta \Delta^2 (\beta) {\rm
e}^{-\beta^2/2 } \sigma_{\vec k}(\beta) \over 
\int d \beta \Delta^2 (\beta) {\rm e}^{- \beta^2 /
2 } },
\end{equation}
where we have denoted
\begin{equation}
\label{vander}
\Delta^2 (\beta)= \prod_{\alpha>0} (\beta\cdot \alpha)^2.
\end{equation}
Notice that, when we write $\beta$ in terms of the orthogonal basis
  (\ref{ubasis}) or (\ref{obasis}), (\ref{vander}) is indeed the 
square of the Vandermonde determinant in the variables $\beta_j$ (for
  $U(N)$) or $\beta_j^2$ (for $O(2r)$). Therefore, as we anticipated
  before, the asymptotic expansion of the integral is an expansion around 
the corresponding Gaussian ensemble, and the perturbative corrections can
  be evaluated systematically as averages in this ensemble.
 
  We will denote
\begin{equation}
\label{asin}
Z_0=\int d \beta \Delta^2 (\beta) {\rm e}^{-\beta^2/2},
\end{equation} 
so that the partition function on Seifert spaces can be written, 
using (\ref{fians}), 
as
\begin{equation}
\label{pertseries}
\log {Z_k(M) \over Z_{1-{\rm loop}}}=-{1\over 24}dy \phi x + 
\log \biggl( 1+ \sum_{\ell=1}^{\infty} \Bigl( \sum_{\vec k|\sum_j jk_j=\ell} 
{1 \over {\vec k}!} a_{\vec k}^{(c)} {\cal R}_{\vec k} (G)\Bigr) \hat x^{\ell} \biggr).
\end{equation}
In this equation $Z_{1-{\rm loop}}$ is given by
\begin{equation}
\label{onelex}
Z_{1-{\rm loop}}={(-1)^{|\Delta_+|} \over | {\cal W}|\, (2 \pi i )^r} 
\Biggl( { {\rm Vol}\, \Lambda_{\rm w} \over {\rm Vol}\,  \Lambda_{\rm r}} 
\Biggr){{\rm e}^{{ \pi i d \over 4} {\rm sign}(H/P)} \over |P|^{d/2}} 
 Z_0 \, \hat x^{d/2},
\end{equation}
and indeed gives the one-loop contribution around the trivial connection. 
This follows by comparing the exact result with the perturbative expansion
\begin{equation}
\log {Z_k(M) \over Z_{1-{\rm loop}}}= 
\sum_{\ell=1}^{\infty} \biggl( \sum_{\Gamma \in {\cal A}(\emptyset)^{\rm
conn}_{\ell}} 
r_{\Gamma}(G) I_{\Gamma}(M)\biggr)x^{\ell}.
\end{equation}
We also see that, by
comparing (\ref{asin}) and (\ref{pertseries}), we 
can extract the value of the universal perturbative invariants
$I_{\Gamma}(M)$ at each order $x^{\ell}$. In order to do that we just
 have to evaluate ${\cal R}_{\vec
k} (G)$ for all vectors $\vec k$ with $\sum_j j k_j \le \ell$, and also 
the group factors $r_{\Gamma}(G)$ for graphs $\Gamma$ with $2 \ell$ vertices. 
Of course, from a mathematical point of view it is not
obvious that the asymptotic expansion of the exact partition function has
the structure predicted by the perturbation theory analysis. The fact that 
this is the case provides an important consistency check of the procedure. 

\subsection{Evaluating the integrals}

We now address the problem of computing the integrals in
(\ref{matint}). As we explained in section 4, the partition function of
Chern-Simons theory on Seifert spaces 
can be interpreted as a matrix model with an
interaction between eigenvalues of the form $\log (\sinh (\beta_i -
\beta_j))$. In the perturbative approach we have to expand the sin in 
power series, and the integrals ${\cal R}_{\vec k}(G)$ are nothing but
averages of symmetric 
polynomials in the eigenvalues in a Gaussian matrix model. 
We will present two methods 
to compute these averages. 

The first method gives the complete answer only up to $\ell=5$, but it has
the advantage of providing general expressions for any simply-laced gauge
group. The starting point is the following identity:
\begin{eqnarray}
\label{groupid}
& & \int d \beta \, {\rm e}^{-{a\over
2}\beta^2 } \prod_{\alpha>0} 4 \sinh \Bigl( {t (\beta\cdot \alpha) \over 2} 
\Bigr) \sinh \Bigl( {s (\beta\cdot \alpha) \over 2} 
\Bigr)\nonumber\\
&=&\Bigl( { 2 \pi \over a}\Bigr)^{r/2}|{\cal W}| ({\rm det}(C))^{1\over 2}
 {\rm e}^{ {t^2 + s^2 \over 2a} \rho^2}\prod_{\alpha>0} 2 \sinh \Bigl(
{ts (\rho\cdot \alpha) \over 2a} 
\Bigr),
\end{eqnarray}
where $C$ is the Cartan matrix of the group. 
This formula is easily proved by using (\ref{wden}). 
Another useful fact is that $\sigma_1 (\beta)$ can be 
written as (see \cite{dif}, pp. 519-20)
\begin{equation}
\label{quadra}
\sum_{\alpha>0} (\beta \cdot \alpha)^2 = y \beta^2.
\end{equation}
One can easily show that, by expanding (\ref{groupid}) in $s,t$, and by
using (\ref{quadra}), it is possible to determine the integrals 
${\cal R}_{\vec k}(G)$ 
for any gauge group up to $\ell=5$, therefore this is enough for the
computational purposes of the present paper. The answer is given in
terms of $y$, $d$, and the quantities 
\begin{equation}
\label{akas}
\alpha_k =\prod_{\alpha>0} (\alpha \cdot \rho)^{2k}.
\end{equation}
For example, one finds:
\begin{equation}
\label{sample}
{\cal R}_{(0,1,0,\cdots)}(G)=5 d y^2.
\end{equation}
The answers obtained by this method are listed in the Appendix. 

In order to evaluate the integrals (\ref{matint}) for arbitrary
$\sigma_{\vec k}$,
it is important to have a more general and systematic method. 
Here is where the connection to matrix integrals becomes
computationally useful. It is easy to see 
that, since the integrals ${\cal R}_{\vec k}(G)$ 
are normalized, one can evaluate them in $U(N)$ and $O(2r)$ instead of
$SU(N)$ and $SO(2r)$. Therefore, one has
\begin{equation}
\label{gue}
{\cal R}_{\vec k}(SU(N))= 
{1 \over Z_0} \int d\beta \, {\rm e}^{-\sum_j \beta_j^2/2}\prod_{i<j} 
(\beta_i -\beta_j)^2 
\sigma_{\vec k}(\beta),
\end{equation}
where 
\begin{equation}
\label{sigU}
\sigma_n (\beta)= \sum_{i<j}(\beta_i-\beta_j)^{2n}.
\end{equation}
In (\ref{gue}) one integrates over $N$ independent variables $\beta_1,
\cdots, \beta_N$. 
It is clear that the $\sigma_{\vec k}(\beta)$ are symmetric polynomials in
these $N$ variables. One can for example 
write (\ref{sigU}) in
terms of power sum polynomials $P_j (\beta)$ (defined in (\ref{pos})) 
as follows, 
\begin{equation}
\label{sigpis}
\sigma_n (\beta)= NP_{2n}(\beta) + {1 \over 2} \sum_{s=1}^{2n-1} 
(-1)^s {2n \choose
s} P_s(\beta) P_{2n-s}(\beta).
\end{equation}
The averages of symmetric polynomials in the Gaussian unitary
ensemble can be evaluated in principle 
by using the Selberg integral \cite{mehta}, or the 
results of \cite{bh}. A more effective way is the
following: any symmetric polynomial in the $\beta_i$'s can be written as a
linear combination of Schur polynomials $S_{\lambda}(\beta)$, which
are labeled by Young tableaux associated to a partition $\lambda$ 
(see (\ref{schur})). Therefore, if we know how to compute the normalized 
average of a Schur polynomial,
\begin{equation}
\label{averdef}
\langle S_{\lambda}(\beta) \rangle= 
{1 \over Z_0} \int d\beta \, {\rm e}^{-\sum_j \beta_j^2/2}\prod_{i<j} 
(\beta_i -\beta_j)^2 
S_{\lambda}(\beta) ,
\end{equation}
we can compute all ${\cal R}_{\vec k}$.  
An explicit expression for (\ref{averdef}) has been
presented in \cite{dfi}. The result is the following: let 
$|\lambda|$ be the total number of boxes in the tableau labeled by
$\lambda$, and let $\lambda_i$ denote the number of boxes in the $i$-th row of
the Young tableau. Define now 
the $|\lambda|$ integers $f_i$ as follows
\begin{equation}
\label{fis}
f_i=\lambda_i +|\lambda| -i,\,\,\,\,\,\, i=1, \cdots, |\lambda|.
\end{equation}
Following \cite{dfi}, we will say that the Young tableau associated to 
$\lambda$ 
is even if the number of odd $f_i$'s is
the same as the number of even $f_i$'s. Otherwise, we will say that it is
odd. If $\lambda$ is odd, the normalized average 
$\langle S_{\lambda}(\beta) \rangle$ vanishes. Otherwise, it is given by:
\begin{equation}
\label{averun}
\langle S_{\lambda}(\beta) \rangle = (-1)^{A(A-1)\over 2} 
{\prod_{f \, {\rm odd}} f!! \prod_{f' \, {\rm even}} f'!! 
\over \prod_{f \, {\rm odd}, f'\, {\rm even}} (f-f')} {\rm dim}\, \lambda,
\end{equation}
where $A=\ell/2$ (notice that $\ell$ has to be even in order to have 
a non vanishing result). Here ${\rm dim}\, \lambda$ is the dimension of the
irreducible representation of $SU(N)$ associated to $\lambda$, 
and can be computed by using the hook formula. This expression solves the 
problem of computing the averages (\ref{gue}) in the general case: we
express the product of power sums appearing in (\ref{sigpis}) in terms of
Schur polynomials by using
Frobenius formula (\ref{frob}), and then we compute the averages of these 
with (\ref{averun}). As
an example of this procedure, let us compute ${\cal
R}_{(0,1,0,\cdots)}(SU(N))$. Using (\ref{sigpis}) and 
Frobenius formula (\ref{frob}), we find:
\begin{equation}
\sigma_2 = (N-1) S_{\tableau{4}} -(N+1) S_{\tableau{1 1 1 1}} + 
(N-3) S_{\tableau{2 1 1}} -(N+3) S_{\tableau{3 1}} + 10 S_{\tableau{2 2}}.
\end{equation}
The averages of the different Schur polynomials can be computed from
(\ref{averun}), and we obtain, after some simple algebra:
\begin{equation}
{\cal
R}_{(0,1,0,\cdots)}(SU(N))=5 N^2 (N^2-1),
\end{equation}
in agreement with (\ref{sample}).

Let us now consider the orthogonal ensemble. The averages that we want to
compute are given by 
\begin{equation}
\label{goe}
{1 \over Z_0} \int d\beta \,  {\rm e}^{-\sum_{j=1}^r \beta_j^2/2}\prod_{1 \le i<j\le
r} 
(\beta^2_i -\beta^2_j)^2 
\sigma_{\vec k}(\beta),
\end{equation}
where 
\begin{eqnarray}
\label{goesig}
\sigma_n (\beta)& = & 
\sum_{i<j} \Bigl\{ (\beta_i + \beta_j)^{2n} +
 (\beta_i - \beta_j)^{2n} \Bigr\} \nonumber\\
&=& (2r- 2^{2n-1})P_n (\beta_i^2)+ \sum_{s=1}^{n-1}{2n \choose 2s} 
P_s (\beta_i^2)  
P_{n-s}(\beta_i^2).
\end{eqnarray}
The 
functions $\sigma_{\vec k}(\beta)$ are now symmetric polynomials in the 
$\beta_i^2$, so we can write them in terms of Schur polynomials 
$S_{\lambda}(\beta_i^2)$. This allows to express the integrals 
(\ref{goe}) in terms of the integrals
\begin{equation}
\int_0^{\infty} \cdots \int_0^{\infty}\,  dy \, \Delta^2(y)
(y_1 \cdots y_r)^{\alpha-1}{\rm e}^{-(y_1+ \cdots+ y_r)/2} S_{\lambda}(y),
\end{equation}
which are a special case of a generalization of the
Selberg integral studied by Kadell
\cite{kadell}, see also \cite{macdon}. Their value is given by
\begin{equation}
r! \prod_{i=1}^r \Gamma (\lambda_i + \alpha + r-i) 
\prod_{i<j} (\lambda_i -\lambda_j + j-i).
\end{equation}
In our case, $\alpha=1/2$. The normalized average of a Schur polynomial 
is then:
\begin{equation}
\label{oaver} 
 \langle S_{\lambda}(\beta_i^2) \rangle =2^{|\lambda|}
{\rm dim} \, \lambda \prod_{i=1}^r { 
\Gamma (\lambda_i + 1/2 + r-i) \over \Gamma ( 1/2 + r-i)}
\end{equation}
In this equation, ${\rm dim}\, \lambda$ denotes 
the dimension of the representation of
$SU(r)$ associated to $\lambda$. This solves the problem of computing 
the averages (\ref{goe}) in the orthogonal ensemble. As a simple
example, let us consider again $\vec k =(0, 1,0, \cdots)$. 
It is easy to see that 
\begin{equation}
\sigma_2=(2r-2) S_{\tableau{2}} + (14-2r)S_{\tableau{1 1}},
\end{equation}
and one finds
\begin{equation}
{\cal R}_{(0, 1,0, \cdots)}(SO(2r))=20 \, r(2r-1)(r-1)^2,
\end{equation}
in agreement with (\ref{sample}).     

\subsection{Universal perturbative invariants up to order 5}

Using the above ingredients, it is easy to find the universal perturbative
invariants of Seifert spaces up to order 5. Although the coefficients $a_s$
in (\ref{integrand}) are functions of $n$ and the Newton polynomials
$P_{\vec k}(p_i^{-2})$, the answer turns out to be 
more compact when written in terms of elementary symmetric
polynomials $E_k$ in the variables $p_i^{-2}-1$ by using (\ref{elpos}) 
(so for example $E_1=-n + \sum_{i=1}^n p_i^{-2}$). One finds, 
\begin{eqnarray}
I_{\theta}&=&-{1 \over 48}\Bigl( \phi - {P \over H}(2 + E_1) \Bigr), 
\nonumber\\ 
I_{\theta_2}& =& {1 \over 1152} \Bigl( {P \over H}\Bigr)^2 (1 + E_1 + E_2), 
\nonumber\\
I_{\theta_3}&=& {1 \over 13824} \Bigl({P \over H}\Bigr)^3 E_3, \nonumber\\
I_{\theta_4}&=& {1 \over 11059200} \Bigl({P \over H}\Bigr)^4 
\Bigl(82 E_4 -46 E_3 -18(1+ E_2 + E_3)E_1 - 9 E_2^2 - 18 E_2 - 9 E_1^2 -9 
\Bigr), \nonumber\\
I_{\omega}&=& {1 \over 1382400}\Bigl({P \over H}\Bigr)^4 \Bigl(2 E_4 - 6
E_3  + 2E_1 (1 + E_2 -E_3) + E_2^2 + 2E_2 + E_1^2 +1 \Bigr),\nonumber\\
I_{\theta_5}&=&{1 \over 66355200}\Bigl({P \over H}\Bigr)^5 \Bigl( 55E_5 +
E_4 ( 27 E_1 -56) - E_3 (9E_2 + 36 E_1 +8)\Bigr), \nonumber\\
I_{\omega\theta}&=&{1 \over 8294400}\Bigl({P \over H}\Bigr)^5 \Bigl( 5 E_5 
-E_4 (3 E_1 +16) + E_3( E_2 + 4 E_1 + 12) \Bigr).
\end{eqnarray} 
Note that the first universal invariant is given by
\begin{equation}
I_{\theta}= {\lambda (M) \over 2},
\end{equation}
where $\lambda (M)$ is the Casson
invariant of $M$, in accord with the general result of \cite{roz} and 
with the result for the LMO
invariant \cite{lmo}. We have also checked that the value for 
$I_{\theta_2}$ listed above agrees with the value
obtained in \cite{bl} for the LMO invariant using the Aarhus integral. 
It would be interesting to see if these invariants have some nice
integrality properties. The perturbative $SU(2)$ invariants of integral
homology spheres do exhibit some
integrality properties (discussed for example in \cite{lr}), but they
include extra factors coming from the $SU(2)$ group weights. In general,
the invariants listed above are rational even for integral homology
spheres. For example, $I_{\theta_3} \in {\bf Z}/4$ for Brieskorn integral
homology spheres. Also, $I_{\theta_2} -{1\over 1152} \in {\bf Z}/2$ for those
spaces. 

\section{Open problems}

Besides the original motivation of understanding universal perturbative
invariants from a field theory point of view, the results presented here 
also provide a computationally feasible
framework to study the Chern-Simons partition function with higher rank
gauge groups. There are various avenues to explore in the context of
Chern-Simons theory and the theory of three-manifold invariants. It would be
interesting for example to understand the structure of perturbation theory in 
the background of a
reducible nontrivial connection, and work out the asymptotic expansion
starting from (\ref{fians}). One could also 
consider other three-manifolds (not necessarily 
rational homology spheres) and see in particular if the matrix model
representation provided here can be generalized to other cases.
      
There are also various interesting physical contexts in which the results of
this paper might be relevant. Let us end by mentioning a few of them:

1) The computation of Rozansky-Witten invariants 
   \cite{rw} 
   involves the universal perturbative invariants of 
three-manifolds that are extracted from Chern-Simons theory, but the weight
   system is now associated to a hyperK\"ahler manifold (see
   \cite{sawth} for a very nice review). Therefore, the universal
   perturbative invariants of Chern-Simons theory are relevant in 
Rozansky-Witten theory. On the other hand, this theory is an essential
   ingredient in 
    the worldvolume theory of M2 membranes in manifolds of $G_2$ holonomy
   \cite{hm}, and the computation of the Rozansky-Witten partition function
   should play a role in understanding membrane instanton effects in $G_2$
   compactifications of M theory.

2) Chern-Simons theory on a three-manifold $M$ describes topological
   A branes wrapping the Lagrangian submanifold $M$ in the Calabi-Yau
   $T^*M$ \cite{wittenopen}. Moreover, the perturbative invariants of
   Chern-Simons theory correspond to topological open string amplitudes on that
   target. Having a systematic procedure to compute Chern-Simons 
  perturbative invariants may prove to be useful in further understanding
   topological open strings in those backgrounds. 

3) Another consequence of our results is that topological 
A branes in $T^*M$ are
   described by a matrix model, when $M$ is a Seifert sphere. It has been
recently shown \cite{dv} that topological B branes on some noncompact
   Calabi-Yau spaces are described by a Hermitian matrix model
characterized by a potential $W(\Phi)$ with multiple cuts. 
It would be interesting to know if there is a relation between the matrix
models of \cite{dv} and the ones presented here. Notice
   that according to our results the partition function of $U(N)$
   Chern-Simons theory on ${\bf S}^3$ can be written as
\begin{equation}
\label{uncs}
Z={ {\rm e}^{- { x \over 12}N(N^2-1)}\over N!} 
\int\prod_{i=1}^N {d \beta_i \over 2\pi} \, {\rm e}^{-\sum_i \beta^2_i/2 x} \prod_{i<j} \Bigl( 2 \sinh {\beta_i - \beta_j\over 2} \Bigr)^2,
\end{equation}  
This describes open topological A strings on $T^* {\bf S}^3$ with $N$ branes
wrapping ${\bf S}^3$, or equivalently
(after the geometric transition of \cite{gvcs}) 
closed topological strings on the resolved
conifold. In the limit $ x \rightarrow
0$, (\ref{uncs}) gives the standard Gaussian model, as 
we have argued at length in this paper. On the other hand, it is shown in
\cite{dv} that the Gaussian model (which corresponds to $W(\Phi)=\Phi^2$) 
describes type topological B strings in the
 deformed conifold 
geometry. This is consistent with the
fact that, as explained in \cite{vafa}, the deformed conifold geometry
gives the mirror of the resolved conifold only at small 't Hooft coupling 
$t=Nx$,
which for fixed $N$ means precisely small $x$. This also suggests to
consider multicut matrix models with a potential $W(\Phi)$, as in \cite{dv},
but where the eigenvalue interaction is not the usual one 
$\prod_{i<j} (\beta_i-\beta_j)^2$ but $\prod_{i<j}(2 \sinh
((\beta_i-\beta_j)/2))^2$. In view of the above observation, these deformed
models might be relevant to understand the mirror of the geometric  
transition studied in \cite{cachazo,dv}. Indeed, 
a compact version of this, corresponding to a unitary 
matrix model where the $\beta_i$ are 
periodic variables, has been considered in \cite{dvtwo} in order 
to describe the stringy realization of the ${\cal N}=2$ Seiberg-Witten 
geometry.
   
4)  Although in this paper the focus has been on the perturbative
   expansion of the partition function, 
the matrix model is also very useful to understand its large
   $N$ expansion. This is an interesting problem in itself, and it 
would be nice to see what is the connection
   to the approach of \cite{douglas}. But of course the large $N$ expansion of
   these models is particularly
   interesting in view of the large $N$ dualities involving Chern-Simons
   theory \cite{gvone,gvcs}. Although these dualities are not expected to
   hold for arbitrary Seifert spaces, the results presented here 
may be useful to understand in detail the case of lens spaces (already
   analyzed in \cite{gvone}) and shed light on the situation for
 more general three-manifolds.

\section*{Acknowledgements}
I would like to thank Bobby Acharya, Mina Aganagic, Emanuel Diaconescu,
Jaume Gomis, Rajesh Gopakumar,
Shinobu Hikami, Jose
Labastida, Greg
Moore, Boris Pioline, Justin Sawon, Toshie Takata, 
Miguel Tierz and Cumrun Vafa for useful discussions
and correspondence. Thanks to Justin too for providing me the graphs of his
thesis. This work is supported by the grant 
NSF--PHY/98--02709.

\appendix

\section{Appendix}

\subsection{Group theory factors}

We first present the group theory factors associated to the connected graphs 
that give a basis of ${\cal A}(\emptyset)^{\rm conn}$ up to order 5. The
evaluation of these factors is straightforward by using the
graphical techniques of Cvitanovi\'c \cite{cvita}, and rather immediate for
all of them (except for 
$r_{\omega}(G)$, that gives 
the quartic Casimir in the adjoint and has been computed for all gauge
groups in the second reference of
\cite{cvita}). Our conventions are as in \cite{cvita}: 
the Lie algebra in the defining
representation has Hermitian generators $T_i$, $i=1, \cdots, d$ satisfying
the commutation relations $[T_i, T_j]=iC_{ijk}$. The generators are
normalized in such a way that the quadratic Casimir of the adjoint
representation $C_A$ (which is defined here by $C_A \delta_{ij}=
\sum_{k,l}C_{ikl}C_{jkl}$) is twice the dual Coxeter
number. This implies that ${\rm Tr}(T_i T_j)= a \delta_{ij}$ with $a=1$ for
$SU(N)$ and $Sp(N)$, and $a=2$ for $SO(N)$ (notice that 
these normalizations differ from the
ones in \cite{al,alp}). 
 
\renewcommand{\arraystretch}{1.50}
\TABULAR{|c|c|c|} {
 \hline
  & $SU(N)$ & $SO(N)$  \\  \hline
 $d$ & $N^2-1$ & ${1\over 2}N(N-1)$   \\ 
\hline
$y$ & $N$ & $N-2$ \\ \hline
$\alpha_2$ & ${1 \over 60} N^2 (N^2-1)(2 N^2-3)$ & ${1 \over 480} 
N(N-1)(N-2) (8 N^3 -45 N^2 + 54 N + 32)$ \\ \hline} 
{\label{grouptable}Dimensions, dual Coxeter numbers and $\alpha_2$ for 
$SU(N)$ and $SO(N)$}

The group factor will be written in terms of the 
dual Coxeter $y$, the dimension of the group $d$, 
and $\alpha_2$ (where $\alpha_k$
is defined in (\ref{akas})). Their values for $SU(N)$, $SO(N)$ are listed in
Table \ref{grouptable}. The results for $Sp(N)$ follow from the
$Sp(N)=SO(-N)$ relation \cite{ck}, so for $Sp(N)$ one has $d=N(N+1)/2$,
$y=N+2$ and $\alpha_2^{Sp(N)}(N)=\alpha_2^{SO(N)}(-N)$. 
The group theory factors for the 
graphs in (\ref{graphi}) are listed in Table \ref{groupfac}.

\begin{table}[htbp] 
\centering 
\begin{tabular}{|c|c|c|} 
 \hline
 $\ell$ & graph & group factor  \\  \hline
 $1$ & $\twoVgraph$  & $2 dy$\\ \hline
$2$ & $\fourVgraph$ & $4 d y^2 $ \\ \hline
$3$ & $\sixVgraph$ &  $8 dy^3 $ \\ \hline
$4$ & $\eightVgraphI$ & $16 d y^4 $ \\ 
 & $\eightVgraphII$ & $18  d y^4 - 480 \alpha_2$ \\ \hline
$5$ & $\tenVgraphI$ & $ 32 d y^5 $ \\
 & $\tenVgraphII$ &  $2 y (18 d y^4 - 480 \alpha_2)$ \\ \hline
\end{tabular}
\caption{Group theory factors for the Feynman graphs up to
 $\ell=5$.}
\label{groupfac}
\end{table}

\subsection{Matrix integrals}

We now list the results for the matrix integrals (\ref{matint}), up to
order 5. The results for $k_1=0$ are:
\begin{eqnarray}
\label{matres}
{\cal R}_{(0,1,0, \cdots)}(G)&=& 5\, dy^2, \nonumber\\
{\cal R}_{(0,0,1,0, \cdots)}(G)&=& 35\, d y^3 , \nonumber\\
{\cal R}_{(0,2,0,0, \cdots)}(G)&=& 25\, d(d+12)y^4 -2880 \alpha_2, 
\nonumber\\
{\cal R}_{(0,0,0,1,0, \cdots)}(G)&=& 350\, d y^4 -1680 \alpha_2, 
\nonumber\\
{\cal R}_{(0,1,1,0,0, \cdots)}(G)&=& 35\, y \bigl\{ 5 d(d+24) y^4 -1728
\alpha_2  \bigr\}, 
\nonumber\\
{\cal R}_{(0,0,0,0,1,0, \cdots)}(G)&=& 4620\,  y  \bigl\{ d y^4 -12
\alpha_2 \bigr\}. 
\end{eqnarray}
The results for $k_1>0$ can be obtained from (\ref{matres}) very easily: 
insertions of $\sigma_1 (\beta)$ can be reduced to
insertions of $\beta^2$ by using (\ref{quadra}), and these can be computed 
by taking derivatives with respect to
$a$ in  (\ref{groupid}). We have for example: 
\begin{equation}
{\cal R}_{(2,1,0, \cdots)}(G)= 5 d (d+4) (d+6) y^4.
\end{equation}
Finally, the integral $Z_0$ in (\ref{asin}) is given by
\begin{equation}
\label{cenot}
Z_0= (2 \pi)^{r\over 2}|{\cal W}| ({\rm det}\,C)^{1 \over 2} \prod_{\alpha
>0} (\alpha \cdot \rho).  
\end{equation} 
Note that this combines with the rest of the factors in (\ref{onelex}) to
produce, up to an overall phase 
\begin{equation}
\label{exloop}
Z_{1-{\rm loop}}= 
{ (2 \pi)^{|\Delta_+|} \over (l|H|)^{d/2}} 
\Biggl( { {\rm Vol}\, \Lambda_{\rm w} \over {\rm Vol}\,  \Lambda_{\rm r}} 
\Biggr)^{1\over 2} \prod_{\alpha
>0} (\alpha \cdot \rho), 
\end{equation}
where we have used that ${\rm Vol}\, \Lambda_{\rm r} / 
{\rm Vol}\,  \Lambda_{\rm w}={\rm det}(C)$. 
If instead of taking the volume of the root lattice in (\ref{exloop}) we
take that of the coroot lattice, the resulting 
expression for the partition function is probably valid for any gauge
group.

\subsection{Symmetric polynomials}
Here we summarize some ingredients of the elementary 
theory of symmetric functions that are used in the paper. A standard
reference is \cite{macdon}. 

Let $x_1, \cdots, x_N$ denote a set of $N$ variables. The {\it elementary
symmetric polynomials} in these variables, $E_m (x)$, are defined as:
\begin{equation}
\label{el}
E_m (x) =\sum_{i_1< \cdots <i_m} x_{i_1} \cdots x_{i_m}.
\end{equation}
The products of elementary symmetric polynomials provide a basis for the 
symmetric functions of $N$
variables with integer coefficients. Another basis is given by the {\it 
Schur polynomials}, $S_{\lambda}(x)$, which are labeled by Young
tableaux. A tableau will be denoted here by a partition $\lambda= 
(\lambda_1, \lambda_2, \cdots, \lambda_p)$, where $\lambda_i$ is the number
of boxes of the $i$-th row of the tableau, and we have 
$\lambda_1 \ge \lambda_2
\ge \cdots \ge \lambda_p$. The total number of boxes of a tableau will be
denoted by $|\lambda|=\sum_i \lambda_i$.
The Schur polynomials are defined as quotients of determinants, 
\begin{equation}
\label{schur}
S_{\lambda}(x)={ {\rm det}\, x_j^{\lambda_i + N-i} \over 
{\rm det}\,  x_j^{ N-i} }.
\end{equation}
A third set of symmetric functions is given by the {\it Newton
polynomials} $P_{\vec k}(x)$. These are labeled by vectors 
$\vec k=(k_1, k_2, \cdots, k_p)$, where the $k_j$ are nonnegative 
integers, and they are defined as
\begin{equation}
\label{newt}
P_{\vec k}(x) = \prod_{j=1}^p P_j^{k_j}(x) ,
\end{equation}
where
\begin{equation}
\label{pos}
P_j(x) =\sum_{i=1}^N x_i^j,
\end{equation}
are power sums. The Newton polynomials are homogeneous of degree
$\ell=\sum_j jk_j$ and give a basis for the symmetric functions 
in $x_1, \cdots, x_N$ with rational coefficients. They 
are related to the Schur polynomials
through the Frobenius formula,
\begin{equation}
\label{frob}
P_{\vec k}(x)= \sum_{\lambda} \chi_{\lambda}(\vec k) S_{\lambda}(x),
\end{equation}
where the sum is over all tableaux such that $|\lambda| = \ell$, 
and $\chi_{\lambda}(\vec k)$ is the character of the symmetric group in
the representation associated to $\lambda$ and evaluated on the conjugacy class
associated to $\vec k$ (this is the conjugacy class with $k_j$ cycles of
length $j$). 
Finally, one has the following relation between elementary symmetric
polynomials and Newton polynomials,
\begin{equation}
\label{elpos}
E_m (x) =\sum_{\vec k} {(-1)^{\sum_j (k_j-1)} \over \prod_j k_j! j^{k_j}} 
P_{\vec k}(x),
\end{equation}
where the sum is over all vectors with $\sum_j jk_j =m$.

\bibliographystyle{plain}
%\bibliography{papers}

\end{document}